# Governing for Free: Rule Process Effects on Reddit Moderator Motivations

Hannah M. Wang[1], Beril Bulat[1], Stephen Fujimoto[1], and Seth Frey[1,2]

[1] University of California, Davis, Davis CA 95616, USA
[2] Ostrom Workshop, Indiana University, Bloomington IN 47408, USA
hawang@ucdavis.edu, bbulat@ucdavis.edu,
ssfujimoto@ucdavis.edu, sethfrey@ucdavis.edu

**Abstract.** Developing a strong community requires empowered leadership capable of overcoming governance challenges. New online platforms have given users opportunities to practice governance through content moderation roles. The over 2.8 million "subreddit" communities on Reddit are governed by hundreds of thousands of volunteer moderators, many of whom have no training or prior experience in a governing role. While moderators often devote daily time to community maintenance and cope with the emotional effects of hate comments or disturbing content, Reddit provides no compensation for this position. Thus, moderators' internal motivations and desire to continue filling this role is critical for their community. Drawing upon the relationship between governance procedures and internalized motivation, we investigate how the processes through which subreddit moderators generate community rules increase moderators' motivation through the meeting of social-psychological needs: Procedural Justice and Self Determination, and Self-Other Merging. Preliminary analysis of survey data from 620 moderators across Reddit shows a correlation between moderators' administrative behaviors and the social-psychological needs underpinning their motivations. Understanding these relationships will allow us to empower moderators to build engaging and cooperative online communities.

## 1 Introduction

What do volunteer moderators of online communities get out of moderating, and what does it mean for peer production and online collective action? Platforms such as Reddit have captured the attention of HCI scholars interested in applying frameworks of self-governance and cooperative resource management to a multitude of communities run by amateurs. The over 2 million "subreddit" communities aggregated on Reddit.com are each independently managed by a team of volunteer moderators, who are responsible for the creation of community rules and the sanctioning of members who break those rules. While past scholarship has centered around the development of these communities across time, and the relationship between governance and community maturity [10], the motivations of moderators to continue developing successful communities are relatively under-researched.

Due to the lack of formal compensation provided to moderators by Reddit, Inc., we propose that internal motivating factors — the satisfaction of social-psychological needs — promote moderators' continual commitment to their community. Our key finding was that the level of hierarchical versus participatory governance in a community is associated with different levels of need fulfillment. More hierarchical structures preserve moderators' sense of Procedural Justice and Self Determination, the belief that the community is run fairly and grants members independence. On the other hand, more participatory structures are associated with higher levels of Self-Other Merging, or the extent to which moderators perceive a cohesive community identity. Understanding the importance of these needs, as well as how governance structures help fulfill them, will allow us to design systems that empower and motivate moderators.

## 2   Literature Review

There are a variety of communities online where a group of people connect around a shared purpose and interact with each other by following established norms, rules, and protocols [8, 9, 18]. Online communities present an example of information commons, as they provide collective goods such as social network capital and knowledge capital [19] and distribute digital resources such as information and creative content that are produced collectively [11]. An information common can be described as a "highly accessible, self-rising information system in which stakeholders share an overarching goal" [16]. Online communities, being self-rising, rely on voluntary efforts and continuous provision of resources, and require sustainability to realize the community's shared goal. This reliance, however, also brings forward a dilemma, where increased participation in a community can both facilitate and hinder user attention, a scarce resource for online communities, when the increased activity also includes unwanted user behaviors [12, 20]. Thus, effective governance is critically essential to any online community's success and endurance.

Online communities are self-governing social systems where the responsibility of the governing mechanism often lies with community moderators who monitor and regulate behaviors through formal rules and policies. Moderators voluntarily carry the burden of drafting, implementing, and updating community rules as needed, usually with little to no training [2]. Individuals tasked with community moderation not only invest time and energy but also emotional labor in upholding community values against potential aggressors that violate community norms [5]. Despite the significance of their roles in facilitating cooperation and engagement in online communities, very little is known about the motivations that drive moderators to fulfill their voluntary roles in the first place. Previous work on online community moderators have considered how formal leadership and community characteristics affect individuals' leadership efforts [1], how they engage with and develop their communities [21], how they foster public discussion [13], and their engagement in collective action across communities [14].

However, there is a research gap pertaining to internal motivations of community moderators, and how different forms of community governance affect these motivations.

Past scholarship in psychology points to the importance of internal motivation for success in any endeavor. When individuals are motivated by self-fulfillment as opposed to the promise of a reward, they are more enthusiastic about their task and tend to perform it better than their externally motivated counterparts [4]. In a governance setting, citizens only reap the benefits of a participatory structure when the governing body is willing to implement it [24]. Research on local governments highlights the importance of leaders' personal ideologies and motivation for participatory governance in the successful implementation of these structures [23].

More specifically, experimental results from DeCaro and colleagues proposed a set of six social-psychological needs as being strongly associated with internal motivation in the governance of a common-pool resource [3]. The researchers proposed that governance structures promote cooperation through the satisfaction of these needs: Procedural Justice and Self-Determination, Belonging, Competence, Security, Interpersonal Justice, and Self-Other Merging. Using a survey based on established measures, it was found that forms of governance – voting on rules versus having them imposed, enforcement versus lack of enforcement – were associated with different levels of these social-psychological needs, and thus internal motivation.

While DeCaro's experiment looked at social-psychological needs in response to governance methods, he only examined two governance conditions – voting and enforcement [3]. Far truer to the reality of online social systems is a continuum of governance structures from hierarchical (one moderator making all decisions) to participatory (all community members involved in decision-making). A series of structured interviews provides four common processes through which moderators guide their communities through the creation of community rules: one moderator making rules alone, a group of moderators making rules, a small group of community members aiding in rule creation, or the entire community working together to make rules [21]. Furthermore, DeCaro's original experiment did not consider the wealth of community diversity apparent on digital platforms. Communities on Reddit range in complexity from small groups of classmates to worldwide social activism movements, each presenting a unique environment in which moderators may experience different challenges and gratifications.

These results underscore the importance of social-psychological need satisfaction in the development of successful systems of governance, as well as the need to look to additional elements of community context when analyzing these outcomes. Several past studies have researched the impacts of governance systems on online communities, finding that users tend to view democratic systems as more legitimate and/or empowering [6, 7, 17]. But such structural institutional measures disregard moderators' social-psychological need satisfaction, which provide the most micro-level basis for higher-level manifestations of self-governance and institutional support for empowerment. In fact, when interviewed about implementing democratic systems, Reddit moderators were conflicted. While many admire the idea of community participation, moderators question the ability of community members to make wise decisions in the group's best interest [15, 21]. In some cases, moderators felt unsure as to whether to believe in the legitimacy of votes. Since Reddit is a public platform with no way of restricting access to content, votes could easily come from users outside the core community, or users intentionally choosing a less favorable option for humor or mischief [15].

Due to this potential for interference, online community leaders often gravitate to an emergent hierarchical order, in which one or a group of moderators has the final say in community policies or decisions. Findings show that the likely fate of most online communities is one with power concentrated within a small group of leaders who exert control over their communities [22]. However, this is not necessarily a bad thing as some evidence suggests that communities benefit from the control of one or a few strong moderators [9]. In fact, many moderators seem to prefer these hierarchical structures, not necessarily out of megalomania, rather a desire to avoid conflict or chaos. While many researchers are quick to favor democratic forms of governance out of fear for totalitarianism or tyranny, we propose that the "best" form of governance must succeed at preserving moderators' internal motivation via the satisfaction of their social-psychological needs. Such a system promotes their continued willingness to engage in governing on a volunteer basis, regardless of its level of community participation. We propose that there exists a relationship between the level of hierarchy present in a community and the satisfaction of moderators' social-psychological needs, which in turn affects their motivation to continue governing without compensation.

## 3      Current Study

The aim of the present study is to understand how moderators' social-psychological needs are related to the systems of governance they employ in their communities, namely the level of participatory decision making conceptualized via the four structures described above (the "rule processes"). Due to this work's focus on moderators as opposed to group members, we focused primarily on two social-psychological needs in our analysis. First is Procedural Justice and Self-Determination, the level of perceived fairness and influence the rule processes afford moderators as well as the level to which moderators feel as if they can exercise their "true desires." Second is Self-Other Merging, or the level to which moderators believe the rule processes give their community a cohesive identity [3]. We believe that in the context of moderators, the satisfaction of these two needs is the most critical.

Many moderators are guided by a sense of "responsibility" to their community, finding the most gratification when they are helping the community grow and develop [21]. Thus, we believe that this gratification is best found when moderators feel as if they have the power to cultivate their ideal community environment (Procedural Justice and Self Determination) and when they identify strongly with community members (Self-Other Merging). Additionally, we find that other social-psychological needs can be interpreted rather trivially in the context of being a moderator (ex. One would naturally feel secure in a community they created or help to organize). Following the past scholarship on emergent hierarchy, focusing on these two measures provides the psychological prerequisites for community members to support the sustainable operation of a volunteer-run peer production community.

Furthermore, we added community age and size as moderators in this model, in light of the immense diversity in community context present on Reddit. In general, subreddit moderators displayed a desire to avoid conflict within their communities. Moderators reported feelings of stress upon being in an unfamiliar group and encountering challenges such as increases in spam, misunderstandings, or hate speech [15]. Thus, we believe that the relationship between hierarchical governance and Procedural Justice and Self-Determination is stronger in young communities, where a less-established group is more likely to present moderation challenges and require tighter control. Furthermore, we believe that the relationship between participatory governance and Self-Other Merging is stronger in smaller communities, where moderators are more likely to trust a "tight circle" of community members and feel a more cohesive sense of identity. We do not believe that size would affect moderators' Procedural Justice and Self Determination because while a larger group may have more potential to become unruly, its age will truly dictate how familiar moderators are with keeping it under control. Similarly, we believe age will have little effect on Self-Other Merging as moderators will have more challenges in building the needed social capital to establish a cohesive identity in a larger community, regardless of its age.

Thus, we hypothesize the following:

Hypothesis 1: More hierarchical rule processes will be correlated with greater feelings of Procedural Justice and Self Determination among moderators.

Hypothesis 1a: This relationship is moderated by community age, with the relationship being stronger in younger communities.

Hypothesis 2: More democratic rule processes will be correlated with greater feelings of Self-Other Merging among moderators.

Hypothesis 2a: This relationship is moderated by community size, with the relationship being stronger in smaller communities.

## 4 Methods

### 4.1 Participants and Procedure

The survey was hosted on the Qualtrics platform. Participants were directly contacted through Modmail, Reddit's feature for messaging moderators. They were told that the goal of the survey was to "understand the relationship between Reddit use, leadership, and community-building behavior" and were required to provide consent before taking the survey. In addition to core metrics of interest, the survey elicited basic demographic information, questions pertaining to Reddit use and community-building behavior. Participants were incentivized to participate with a small lottery. Upon recruitment, they were also encouraged to share the survey with others. One attention check was contained in the survey; a question asked users to "Please mark 'Strongly Disagree'" and those who did not were omitted. No personally-identifying information was collected or stored.

A total of 2269 respondents took part in the study, however, all but 620 (1649) responses were discarded due to lack of completion, not being a subreddit moderator, or failure to pass the attention check. We suspect that a number of these discarded responses were bots programmed to go through the survey in order to enter the lottery. The 620 remaining participants had a mean age of 31.42 (SD = 10.06), and the majority identified themselves as male (71.09%, n = 450). 45.66% (n = 289) of participants spent less than five hours each week moderating their community. 27.96% (n = 177) spent 5-10 hours, 15.01% (n = 95) spent 10-15 hours, 4.58% (n = 29) spent 15-20 hours, and 6.79% (n = 43) reported spending 20 hours or more.

## 4.2 Measures

**Procedural Justice and Self-Determination (PJSD)**. Four questions in the survey were used to measure the respondent's PJSD score on a 7-point scale. For example: "The procedures used to create rules make me feel as if: I am able to exercise my views and desires." The final PJSD score was the average of the four coded responses (M: 5.89; SD: 1.00; Cronbach's alpha: 0.84).

**Self-Other Merging**. Self-Other Merging score was also calculated in a similar way, with four items, such as "I feel good about the other people in this community." The final Self-Other Merging score was the average of the coded responses (M: 5.50; SD: 1.12; Cronbach's alpha: 0.89).

**Rule Process.** The rule process values were coded responses (1 to 4) from one question asking for the response that best matches the way the community decides on rules. The options for this question ranged from most hierarchical (1) to most democratic (4): 1) moderators create rules on their own; 2) a small group of moderators create the rules; 3) moderators work with a small group of community members; and 4) moderators work with the entire community (M: 2.60; SD: 1.13).

**Community Age.** The survey also asked how old the community is, providing six options (ranging from "less than 6 months" to "10 years or more"); the coded responses to this question (1 to 6 from shortest age to longest age) were used as the values for the community age variable (M: 4.34, corresponding to in between "3-6 years" and "6-9 years"; SD: 1.24).

**Size of Active Community.** The size of the active community variable was coded responses for the question asking for an estimate of the number of community members who make five or more posts or comments per month (ranging from 0 to "more than a thousand"). This measure of activity, i5, has been used elsewhere to determine user engagement on Reddit communities [10]. The responses were coded from 1 to 8 (M: 4.62, corresponding to in between "11-50" and "51-100"; SD: 2.05).

## 5 Results

In order to explore our two research questions, we looked at two models. For Hypothesis 1, the model was made up of rule process as the explanatory variable and PJSD scores as the response variable, with community age as the moderator variable. For Hypothesis 2, the second model was made up of rule process as the explanatory variable and Self-Other Merging scores as the response variable, with size of active community as the moderator variable. The explanatory variable and both moderator variables were centered around the mean of each (the mean was subtracted from each value in the variable). To permit data exploration in a way that mitigated the risk of p-hacking, we pursued a semi-self-replication strategy in which we developed our models on a minority of the data (just 292 observations, 47.1% of the data) before freezing the models and running and reporting them on the full dataset of 620 observations.

### 5.1 Model 1: Procedural Justice and Self-Determination Model

Firstly, the model was significant overall ($F(3, 616) = 9.89$, $p<0.05$). The rule process was negatively associated with PJSD ($\beta=-0.33$, $p<0.05$): moderators in communities with a more democratic rule-making process reported lower PJSD. The interaction between rule process and community age was not significant.

### 5.2 Model 2: Self-Other Merging Model

The second model was also significant overall ($F(3, 616) = 9.51$, $p<0.05$). We found a positive association ($\beta=0.44$, $p<0.05$) between rule process and Self-Other Merging score meaning that the more democratic the rule-making process, the higher Self-Other Merging was reported. We also found a significant negative interaction between rule process and active community size ($\beta=-0.32$, $p<0.05$), pointing to the effect of rule process being larger for smaller communities.

## 6 Discussion

This study explores the reasons why individuals would choose to continue moderating online communities, looking at social-psychological needs in particular as a possible explanation. Using the survey data, the study looked at two hypothesized relationships between rule process and two specific social-psychological needs. The data showed support for most of the two hypotheses.

The first model provided evidence that H1 was correct, showing a significant negative association where the more participatory the rule process reported, the lower the PJSD score was seen overall. This falls in line with the previous findings regarding moderators' preferences for emergent hierarchy. Moderators with more control over their community feel more confident that the procedures used to create the rules are fair because they are protected from users acting outside the community's best interests. The more control they have over their community, the more they feel empowered to act in its best interest. However, the interaction term in Model 1 was not significant and thus H1a could not be supported by the data.

The survey data also showed that democratic rule process was positively correlated with higher Self-Other Merging scores, supporting H2. Such a finding suggests that while they may take away from moderators' sense of justice and fairness, democratic processes do promote a cohesive community identity. This is also important to the social-psychological well-being of moderators, as many align with the philosophy that moderators should act as engaged community members while simultaneously "setting an example" for good behavior [21]. Furthermore, H2a was backed up by the model, with the negative interaction between rule process and active community size being significant. Thus, the sense of cohesiveness that more democratic processes are associated with will likely decrease as the community gets larger.

It may seem provocative to claim that a more hierarchical, less democratic decision structure leads to more "democratic" attitudes of higher Self-Determination, Procedural Justice, and group identity (Self-Other Merging). However, under conditions of an overworked, underappreciated population of volunteer moderators, without whom these communities would not exist, centralization of authority is unarguably pragmatic, and apparently beneficial in such a resource constrained peer production setting.

## 7      Conclusion

Online platforms provide vast arenas for discussion, collaboration, and the creation of shared meaning. However, the unpaid structure of governance underlying such platforms points to the need for internally motivated community moderators. Although much attention has been given to the community implications of governance structures and practices, it is essential to consider their relationships with moderators' social-psychological well-being. Designing systems to best satisfy moderators' social-psychological needs will allow platforms to be sustained by motivated, engaged moderators acting in the best interests of their communities.

## Acknowledgements

The authors wish to thank research assistants Theresa Sims, Jesleyn Gill, Sharon Yoo, Katherine Coviello, Hannah Skepner, Samantha Vigil, Kabir Sahni, Anastacia Dobson Bell, Megan Tsang, Anna Beatrice Ricasata, Sean Abellera, Chengyue Jiang, Evan Brosnan, Yuqi Cheng, Nicole Calbreath, Abigail Endler, Nebiyat Walelign, Harneet Nagra, Alessandra Soto, Mark Murakami, Kelley Ann, Kexin Li, and Yakov Perlov for help with recruitment. Funding for this project was provided by the Provost's Undergraduate Fellowship from the UC Davis Undergraduate Research Center.